\documentclass{elsart}
\usepackage{graphicx}
\begin{document}
\begin{frontmatter}
\title{The importance of nuclear masses in the astrophysical rp-process}

\author{H. Schatz}
\address{Dept. of Physics and Astronomy, National Superconducting Cyclotron 
Laboratory, and Joint Institute for Nuclear Astrophysics, Michigan State 
University, East Lansing, MI 48824, USA}

\begin{abstract}
The importance of mass measurements for astrophysical capture processes in 
general, and for the rp-process in X-ray bursts in particular is discussed. 
A review of the current uncertainties in the effective lifetimes of the 
major waiting points $^{64}$Ge, $^{68}$Se, and $^{72}$Kr demonstrates that 
despite of recent measurements uncertainties are still significant. It is 
found that 
mass measurements with an accuracy of the order of 10~keV or better are desirable,
and that reaction rate uncertainties play a critical role as well. 
\end{abstract}
\end{frontmatter}

\section{Introduction}

Sequences of neutron and proton capture, interspersed with $\beta$ decays, 
play an important role in astrophysics. The 
slow- and rapid neutron capture processes (s- and r-process) are responsible 
for the synthesis of most of the elements beyond the iron group
\cite{Kap99,CTT91}. Slower proton 
captures generate much of the energy in Nova explosions and massive stars, 
while the rapid proton 
capture process (rp-process) powers type I X-ray bursts 
\cite{WaW81,SAG97} and might also 
occur in a proton rich neutrino driven wind in core collapse supernovae
\cite{Fro06,Pru05}.

The nature of such capture processes depends on the temperature and density 
conditions encountered in the stellar environment. At relatively low temperatures and densities
capture reactions are typically much slower than $\beta$
decays. Captures can then only occur on stable or very long lived isotopes 
and the reaction paths proceed along the valley of stability. At somewhat 
higher temperatures and densities encountered mainly in explosive scenarios
capture rates can become faster than 
$\beta$-decay rates and the sequence of reactions responsible for nucleosynthesis and 
energy generation moves towards unstable nuclei. At the most extreme conditions, nucleosynthesis 
paths are governed by partial (QSE, Quasi nuclear statistical equilibrium)
or full nuclear statistical equilibrium (NSE) as both, 
particle capture rates and the 
rates of their inverse photodisintegration processes triggered by high energy photons are fast. 
The nuclei favored in full NSE depend on the conditions, 
in particular electron fraction and entropy, but typically nucleosynthesis paths tend to shift to 
nuclei closer to stability, or, at the extreme involve only
protons, neutrons, and alpha particles. The temperatures and densities during the astrophysical 
rapid neutron capture process (r-process) and rapid proton capture 
processes (rp-process) are just short of establishing NSE. These processes therefore
proceed along some of the most exotic nuclei encountered in astrophysics. Nevertheless,
at these extreme conditions some local equilibrium clusters do form - in the rp-process
along isotonic chains as ($p,\gamma$)-($\gamma,p$) equilibrium,
and in the r-process along isotopic chains as ($n,\gamma$)-($\gamma,n$) equilibrium. These equilibrium 
clusters tend to prevent the reaction paths from reaching or crossing the respective drip lines
and determine the so called "waiting point" nucleus - the nucleus with the 
highest abundance in an equilibrium chain. Once equilibrium is established,
the process has to wait for the waiting point nucleus to $\beta$ decay in order to 
proceed towards heavier nuclei.

In an isotonic or isotopic equilibrium the 
abundance ratio of two neighboring nuclei indexed by $n$ and $n+1$ 
with increasing $Z$ or $N$ is simply given by the Saha equation:

\begin{equation} \label{EqSaha}
\frac{Y_{n+1}}{Y_n} = \rho_n \frac{G_{n+1}}{2G_n} \left(
 \frac{A_{n+1}}{A_n} \frac{2\pi \hbar^2}{m_u kT} \right) ^{3/2}
 \exp \left( \frac{S_{n+1}}{kT} \right)
\end{equation}

where $Y_n$ and $Y_{n+1}$ are the abundances of an initial and final nucleus of a single 
proton or neutron 
capture reaction in the chain, $T$ is the temperature, $\rho_n$ the proton or neutron density, 
$G$ the partition function,
$A$ the mass number, $m_u$ the atomic mass unit, $k$ the Boltzmann constant,
and $S$ the proton or neutron separation energy, respectively.
The maximum abundance
in a chain and therefore the path of the process for a given density and temperature
occurs at a fixed separation 
energy. Because of the exponential dependence on binding energy differences
($S_{n+1}$ in Eq.~\ref{EqSaha})
nuclear masses are among the most important quantities for modeling the r- and rp-processes.

To which degree are equilibria along isotopic or isotonic chains realized in the r- or rp-process? 
Most but not all current r-process models, including models based on
the neutron rich neutrino driven wind in core collapse supernovae, 
are based on a freezeout from some 
high temperature NSE or QSE state and therefore include an extended phase of
more extended ($n,\gamma$)-($\gamma,n$) equilibrium along isotopic chains.
Before the fundamental problem of the site of the r-process is solved, 
it cannot be decided with certainty what the relevant nuclear processes are but to move 
the field forward it is critical to address the nuclear physics issues of the most promising models. 

The situation for the rp-process in neutrino driven winds is similar with the system 
passing through a phase of  ($p,\gamma$)-($\gamma,p$) equilibrium prior to freezeout. On 
the other hand, the rp-process in X-ray bursts is characterized by a rapid heating 
phase (1-10 s) up to peak temperatures around 1.5-2~GK followed by a slower 
cooling phase (10-100 s) and 
freezeout. In X-ray bursts, extended  ($p,\gamma$)-($\gamma,p$) equilibrium in most isotonic 
chains is only established for
the relatively short period during the peak of the burst when temperatures exceed about 1.2-1.3~GK.  
Even then, because of the relatively steep slope of the proton separation energy towards the 
proton drip line, there are still many important reactions where proton separation
energies are too high and ($\gamma,p$) reactions are too slow to establish equilibrium. 
For example, for $N=32$ the 
main rp-process flow proceeds via $^{64}$Ge(p,$\gamma$)$^{65}$As(p,$\gamma$)$^{66}$Se with 
leakages through $\beta$-decays. The proton separation energy of $^{65}$As is low 
(-0.36 MeV, see below) 
and therefore $^{65}$As($\gamma$,p) is fast, establishing   ($p,\gamma$)-($\gamma,p$)
equilibrium during the entire phase of reaction flow during this region. On the other
hand the proton separation energy of $^{66}$Se is 2.4 MeV (see below). 
As we will show below, temperatures of more 
than 1.5~GK are required for $^{66}$Se($\gamma$,p) to establish a  full ($p,\gamma$)-($\gamma,p$) equilibrium 
between $^{65}$As and $^{66}$Se. Therefore, the rp-process in X-ray bursts proceeds through phases of
partial  ($p,\gamma$)-($\gamma,p$) equilibrium, mostly between pairs of isotones
near the proton drip line, and, depending on the peak temperatures reached in the particular
X-ray burst model, a brief phase of complete  ($p,\gamma$)-($\gamma,p$) equilibrium at the 
highest temperatures. 

Therefore, the rp-process in X-ray bursts is a rather complex process. The extent of 
equilibria is rapidly changing within seconds during the burst rise and within 10-100~s 
during the burst cooling. The reaction flows cannot be described simply with 
Eq.~\ref{EqSaha} and, as we will show below, proton capture rates play an important role
during much of the rp-process. In addition, up to about $Z \sim 20$ (depending on the 
peak temperatures attained in a particular X-ray burst) ($\alpha$,p) reactions
can compete with proton capture chains and the respective branchings depend also
on reaction rates. Nevertheless, masses are a critical part of the nuclear physics determining 
observables of X-ray bursts.

The sensitivity of r-process calculations to nuclear masses has been discussed extensively 
in the past (for example \cite{CTT91,PKT01,WGS04}). The rp-process in neutrino driven winds is a rather new 
concept with the added complication of an interplay with neutron induced reactions 
as neutrino interactions do create a sizeable 
neutron density \cite{Fro06,Pru05}.  Pruet et al. 2005 \cite{Pru05} point out that mass uncertainties directly 
affect the final abundances and that improved masses for neutron 
deficient isotopes, for example around $^{92}$Ru, would be important to address the question 
of the contribution of this scenario to galactic nucleosynthesis. 
More work needs to be done to investigate in detail the nuclear physics sensitivities. 
In this paper we therefore concentrate on the role
of masses in the rp-process in X-ray bursts. In contrast to the r-process, a significant 
number of masses along the rp-process have been detemined experimentally. Reviews
of the relevant nuclear physics can be found in \cite{ScR06,SAG97}. As all
but a few rp-process nuclei have been observed in experiments, the 
proton drip line is roughly delineated by experimental constraints on the lifetime 
of nuclei. In addition,
unknown masses can be predicted more reliably as one only needs to extrapolate a few
mass units in most cases. This can be done using the extrapolation method of Audi 
and Wapstra \cite{AME03}. As the rp-process proceeds mostly beyond the $N=Z$ line
one can also take advantage of isospin symmetry and calculate the masses of the most 
exotic rp-process nuclei from the better known masses of their mirrors. Brown et al. 
\cite{AB1} have 
recently shown that mass shifts between isospin mirror nuclei can be calculated with 
an accuracy of 100~keV using a Skyrme Hartree-Fock model. 
However, this still requires accurate knowledge of the masses
of the mirror nuclei that lie closer to stability. Despite of this progress, the typical
theoretical mass uncertainties of many hundreds of keV are still not 
acceptable to reliably model X-ray bursts and to compare calculations with 
observations in a quantitative way. Mass measurements (together
with reaction rate measurements) are therefore essential for 
a better understanding of the rp-process. 

In section \ref{rp-process} we begin by summarizing the astrophysical observables 
that drive the demand for improved nuclear physics in the rp-process. 
After discussing the importance of mass measurements for reaction rate 
calculations, we then focus in section \ref{waitingpoints} on a series of recent precision 
mass measurements performed using ion traps. 
We explore the potential impact on rp-process
calculations and the interplay of masses and reaction rates. In particular we show
that even though tremendous progress has been made through recent experimental 
work, the question of the 
rp-process timescale for passing through the region of the major bottle-necks
$^{64}$Ge, $^{68}$Se, and $^{72}$Kr is still not resolved. 

\section{Masses in the rp-process} \label{rp-process}

The rapid proton capture process powers type I X-ray bursts, which occur when a 
neutron star accretes hydrogen rich matter from a companion star in a binary system. See
\cite{Psa04,StB03} for reviews of the astrophysical aspects 
and \cite{SAG97,ScR06} for a recent review of the nuclear physics aspects. 
The observed burst light curves are sensitive to the underlying nuclear physics 
\cite{AB1,KHA99,SAB01,WHC04}. Nuclear
physics is therefore needed to allow one to interpret burst observations in terms of system 
parameters such as the properties of the neutron star or the composition of the accreted matter. 
In addition, most of the nuclear burning ashes from X-ray bursts remains on the surface of the 
neutron star and gets incorporated into the crust by the ongoing accretion. The nuclear physics
of X-ray bursts is therefore needed to accurately calculate the composition of the 
burst ashes and therefore the composition of the neutron star crust. A wide range of observable crust 
phenomena such as the rare superbursts \cite{ICK04} or the surface cooling behavior of the 
neutron star depend critically 
on composition \cite{Bro04}. This is particularly important as these phenomena could be 
used to constrain the properties of the neutron star if understood in a quantitative way. Finally, it has been 
shown recently that there is the possibility for some types of bursts to eject small but potentially 
observable amounts of burst ashes \cite{WBS06}. Nuclear physics is needed to accurately predict the 
observables and to match possible future observations with X-ray burst models. 

Precision mass measurements are important for rp-process calculations for two reasons. 
First, as explained above,  ($p,\gamma$)-($\gamma,p$) equilibrium clusters form 
during the rp-process with the reaction flow largely determined by mass differences, 
i.e. proton separation energies. The mass sensitivity comes mainly from the 
$\exp(S_{n+1}/kT)$ term in Eq.~\ref{EqSaha}. The second reason is the importance of 
accurate nuclear masses in the calculation of reaction rates, for example for 
proton capture. Most of the relevant nuclear reactions proceed through 
resonances, and the corresponding rate can be expressed as a sum over all 
resonances through \cite{FoH64}

\begin{equation} \label{EqRes}
N_A <\sigma v> = 1.540 \times 10^{11} (\mu T_9)^{-3/2} \sum_j
\omega\gamma_i {\rm e}^{-E_{j}/(kT)} \, \, \, {\rm cm^3 \, s^{-1} mole^{-1}}
\end{equation}

with the resonance energy in the center of mass system $E_{j}$,
the temperature in GK
$T_9$ and the reduced mass
of the entrance channel $\mu$ in amu. The resonance strengths $\omega \gamma_{j}$
are in MeV and can be calculated for proton capture as
\begin{equation} \label{EqOG}
\omega \gamma_{i} = \frac{2J_j+1}{2(2J_T+1)} \frac{\Gamma_{{\rm p}\, j} \Gamma_{\gamma j}}
{\Gamma_{{\rm total}\, j}}
\end{equation}
where $J_T$ is the target spin, and $J_j$, $\Gamma_{{\rm p} \, j}$, $\Gamma_{\gamma \,j}$,
${\Gamma_{{\rm total}\,j}}$
are spin, proton decay width, $\gamma$-decay width, and total width of the compound nucleus state $j$.

Near the proton drip line level densities tend to be low limiting the applicability 
of more reliable statistical model calculations. Direct reaction rate 
measurements using low energy radioactive beams are extremely difficult 
due to limited beam intensities at existing radioactive beam facilities, 
and therefore Eq.~\ref{EqRes}
is often the only way to determine a reaction rate. The various ingredients, such
as particle and radiative widths for the levels involved can be obtained from 
experiments or from shell model calculations \cite{SD1,FP1}. 
However, as Eq.~\ref{EqRes} shows, the reaction rate depends exponentially 
on the resonance energy, which is obtained from the excitation energy $E_x$ of the 
resonant state using the reaction Q-value $Q$ from $E_j = E_x - Q$. 
In addition, the proton width $\Gamma_p$ in Eq.~\ref{EqOG} includes the penetrability of the 
proton through the Coulomb barrier and 
depends therefore also exponentially on the resonance energy.
Reaction rates calculated with Eq.~\ref{EqRes} are therefore extremely 
sensitive to nuclear masses and excitation energies. 
Typically, resonance energies need to be known to 
better than 10~keV to keep the corresponding reaction rate uncertainty below about a 
factor of 2. This is far beyond the accuracy achievable with mass
extrapolations for the Q-values, and shell model calculations for the excitation 
energies. The latter typically reach 100~keV accuracy in the rather well
constrained sd-shell when the levels in mirror nuclei are accurately known
\cite{SBB06}. 

\begin{figure}
\begin{center}
\includegraphics*[width=18cm]{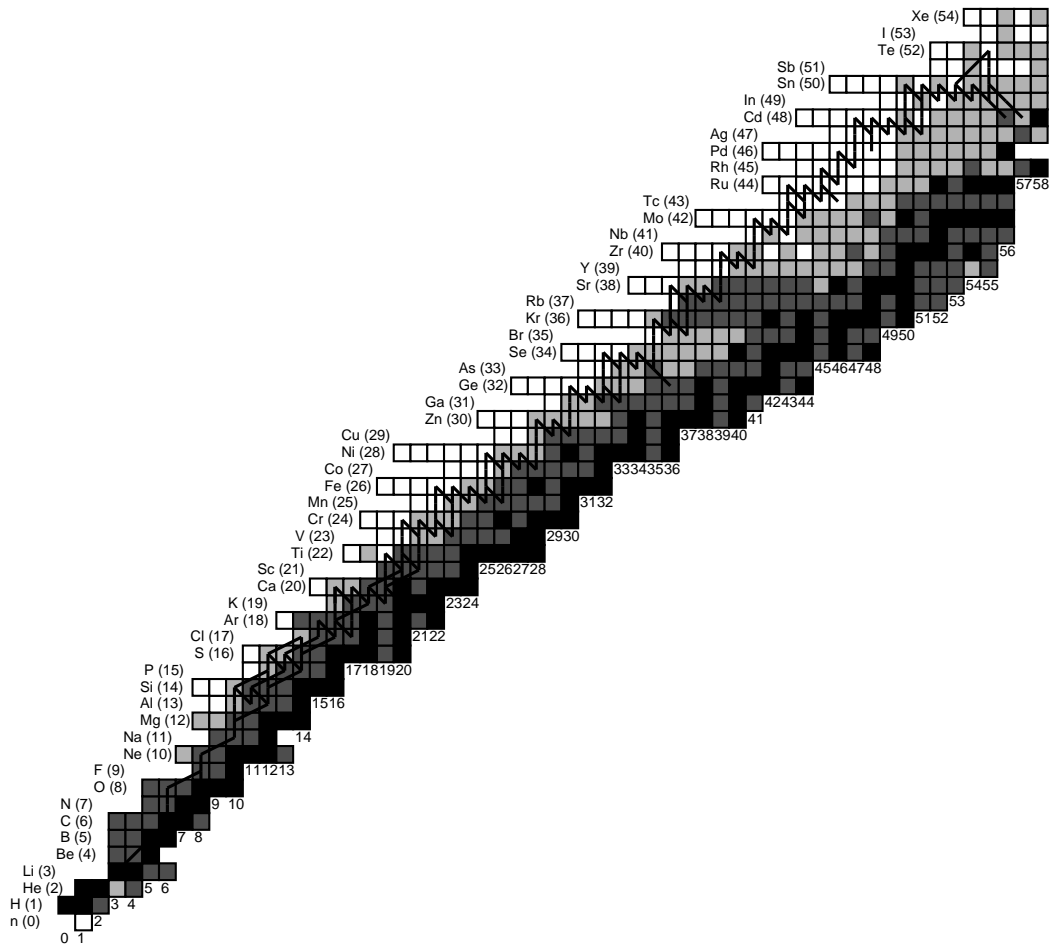}
\end{center}
\caption{The path of the rp-process (from \protect\cite{SAB01}) on the chart of nuclides.
Stable nuclei are black, nuclei with experimentally known masses are grey - dark grey
for uncertainties of less then 10~keV, light grey for larger uncertainties.}
\label{FigMass}
\end{figure}

Recently techniques have been developed to measure excitation energies of 
low lying states in
very neutron deficient nuclei using radioactive beams.
An example is the use of (p,d) transfer reactions in inverse kinematics using 
fast radioactive beams at the NSCL at Michigan State University \cite{SBB06,CLE04}. 
In a first experiment excitation energies in $^{33}$Ar needed for the calculation of 
the $^{32}$Cl(p,$\gamma$)$^{33}$Ar reaction rates could be determined
with accuracies of better than 10~keV. In addition, the mass of 
$^{33}$Ar had been determined independently at ISOLTRAP \cite{BAB03} with 
an accuracy of 0.44~keV.
Combining the mass and excitation energy measurements and using 
shell model calculations for all other level properties the reaction rate 
uncertainty could be reduced from a factor of about 10000 to a factor of 
3-6 \cite{SBB06}. Excitation energy measurements with fast radioactive beams are now 
possible for much of the rp-process up to about A$\sim$68. However,
these measurements need to be complemented with accurate measurements 
of ground state masses of better than 10~keV (1~keV accuracy is desirable) to be useful.
As Fig.~\ref{FigMass} shows masses are currently not known with sufficient accuracy for most 
of the rp-process path,
even when experimental data are available. Further precision measurements of 
ground state masses using ion trap measurements together with measurements of 
excitation energies along the path of the rp-process are therefore needed. 

\section{The $A=64-72$ rp-process bottlenecks} \label{waitingpoints}

A particularly important part of the rp-process is the reaction flow through the region of the 
three major waiting points $^{64}$Ge, $^{68}$Se, and $^{72}$Kr
\cite{SAG97} with $\beta$-decay half-lives $T_{1/2}$ of 63.7~s, 35.5~s, and 
17.16~s \cite{nubase03}, respectively. 
Because these isotopes are located at the proton drip line, 
proton capture is hampered by ($\gamma$,p) reactions on 
the next, most likely proton unbound isotone. If the rp-process had to proceed via the $\beta$-decay of these 
waiting points, their combined $\beta$-decay lifetime $\tau_\beta=T_{1/2}/\ln 2$ of 168~s would represent 
a large impedance for the rp-process that would be comparable to typical burst timescales of 10-100~s. 
This 
would significantly slow down the rate of hydrogen 
burning and lead to extended burst tails. X-ray burst model calculations are therefore particularly sensitive
to the question to which degree weak proton capture flows can reduce these lifetimes. 
This issue is at the heart of the question whether long burst timescales can be used as a signature of the rp-process and if one can use burst timescales to put quantitative constraints on the amount of hydrogen accreted. As discussed and demonstrated in several studies, X-ray 
burst lightcurves are directly sensitive to the effective lifetimes of the 
waiting points $^{64}$Ge, $^{68}$Se, and $^{72}$Kr \cite{SAG97,AB1,KHA99,SAB01,WHC04}.

The nuclear physics determining the effective rp-process lifetime of major waiting 
points has been discussed extensively in Schatz et al. \cite{SAG97}. In the following we discuss in 
detail the situation at the $^{64}$Ge waiting point. Analogeous processes occur at $^{68}$Se and $^{72}$Kr. 
The proton capture flow on $^{64}$Ge proceeds via a so called 2p-capture reaction. 
Current mass predictions (see below) predict $^{65}$As to 
be slightly unbound with a proton separation energy of -0.36 MeV. The Coulomb-barrier is sufficient to suppress spontaneous proton decay so that $^{65}$As decays under terrestrial 
conditions by $\beta$-decay in agreement with experimental evidence \cite{nubase03}. 
The low proton separation energy leads to an establishment of a  ($p,\gamma$)-($\gamma,p$) equilibrium between $^{64}$Ge and $^{65}$As at all times when there is reaction flow in this mass region. Also, because of the low proton separation energy the equilibrium abundance
of $^{65}$As is very low. Nevertheless, proton capture on $^{65}$As can lead to a significant 
2p-capture flow that can reduce the lifetime of $^{64}$Ge during an X-ray burst. At X-ray burst peak conditions $^{64}$Ge, $^{65}$As, and $^{66}$Se are in   ($p,\gamma$)-($\gamma,p$) equilibrium. 
The net 2p-capture flow through this isotonic chain and therefore the effective lifetime of 
$^{64}$Ge is then
determined by the leakage out of the equilibrium via the $\beta$ decay of $^{66}$Se. 
The most critical quantities to 
determine the 2p capture rate and the effective lifetime of $^{64}$Ge for a given temperature
and proton density are therefore the $\beta$-decay half-lives of $^{64}$Ge and $^{66}$Se, the proton separation energies of $^{65}$As and $^{66}$Se, and the proton capture rate on 
$^{65}$As. Accordingly, the $\beta$-decay half-lives of $^{68}$Se, $^{70}$Kr, $^{72}$Kr,
$^{74}$Sr, the proton separation energies of $^{69}$Br, $^{70}$Kr, $^{73}$Rb, $^{74}$Sr,
and the proton capture rates on $^{69}$Br and $^{73}$Rb are of importance for the 
waiting points at $^{68}$Se and $^{72}$Kr respectively. 

While the $\beta$-decay half-lives of the important nuclei around these major waiting points
have been known in most cases 
for some time \cite{nubase03,PrF03}, it has recently become possible to also perform precision mass
measurements using ion traps in this mass region. Of particular importance are recent 
mass measurements of $^{64}$Ge and $^{68}$Se using the Canadian Penning Trap at 
ANL \cite{CLA04,GE64} (the mass of $^{68}$Se had also been determined independently using the $\beta$-endpoint
technique \cite{WAB04}) and the mass measurements of $^{72-74}$Kr at ISOLTRAP \cite{KR72}. 
In addition, Skyrme Hartree-Fock calculations allow now 
the calculation of Coulomb mass shifts between mirror
nuclei with an estimated accuracy of 100~keV \cite{AB1}. While mass measurements 
on $^{65}$As, $^{66}$Se, $^{69}$Br, $^{70}$Kr, $^{73}$Rb, and $^{74}$Sr have not been feasible so far, their isospin mirrors are within reach for ion trap mass measurements. This has 
already been demonstrated for $^{73}$Kr and $^{74}$Kr, which are the mirrors to $^{73}$Rb and
$^{74}$Sr \cite{KR72}. Similar measurements on $^{65}$Ge, $^{66}$Ge, $^{69}$Se and $^{70}$Se would 
be desirable. If an accuracy of 10 keV or better can be reached, these mass measurements can 
be combined with Coulomb shift calculations
to provide mass predictions that are accurate to about 100~keV. 

\begin{table}
\caption{Proton separation energies $S_p$ and uncertainties in MeV used for this study. The masses for
$^{64}$Ge, $^{68}$Se, and $^{72}$Kr are from recent experiments \cite{GE64,CLA04,KR72}, with 
the mass of $^{64}$Ge being taken from a preliminary analysis. The remaining
masses were obtained from the isospin mirrors using the Coulomb shifts from \cite{AB1}. The masses of 
the mirror nuclei $^{73}$Kr and $^{74}$Kr are from \cite{KR72}, for all others from \cite{AME03}.
\label{TabMasses}}
\begin{tabular}{lll} \hline
$S_p(^{65}$As)$=$  -0.36 $\pm$ 0.15 & $S_p(^{69}$Br)$=$  -0.81 $\pm$ 0.10 & $S_p(^{73}$Rb)$=$  -0.70 $\pm$ 0.10 \\
$S_p(^{66}$Se)$=$  2.43 $\pm$ 0.18 & $S_p(^{70}$Kr)$=$  2.58 $\pm$ 0.16 & $S_p(^{74}$Sr)$=$  2.20 $\pm$ 0.14 \\
\hline
\end{tabular}
\end{table}

We use the currently available mass data on rp-process nuclei and their mirrors together with 
Coulomb shift calculations \cite{AB1} to analyze the current uncertainties in the effective 
lifetimes $\tau_{\rm eff}$ of $^{64}$Ge, $^{68}$Se, and $^{72}$Kr with 
$\tau_{\rm eff}=1/(\tau_\beta^{-1} + 
\tau_{2p}^{-1})$, $\tau_\beta$ being the $\beta$-decay lifetime, and $\tau_{2p}$ being 
the lifetime against 2p-capture. The proton separation 
energies used for this study 
are listed in table \ref{TabMasses} with their uncertainties. We follow here the method
outlined in Rodriguez et al. 2004 \cite{KR72} for the case of $^{72}$Kr. For each 
waiting point we performed a small network calculation to determine the effective 
lifetime against proton capture and $\beta$-decay as a function of temperature and 
proton density. The network included the proton captures on the waiting point ($Z,A$) and 
the following isotone, their inverse ($\gamma$,p) reaction rates as well as the 
$\beta$-decay rates of the three nuclei involved. $\beta$-decay rates were taken from nubase \cite{nubase03}
when available. For $^{74}$Sr we used an estimate of 50~ms \cite{PrF03}. For the 
even-even nuclei no changes in half-lives with temperature or density compared to 
terrestrial values are expected for 
X-ray burst conditions \cite{PrF03}. As mentioned below the calculations are insensitive to 
the $\beta$-decay rates of the odd $Z$ nuclei. 
Proton capture rates were taken from the statistical model NON-SMOKER \cite{RaT00}. An upper and
a lower limit within the mass uncertainties was calculated by assuming the upper
one-sigma limits or the lower one-sigma limits for both proton capture Q-values respectively. 
To explore the impact of reaction rates, we in addition increased or decreased the proton 
capture rate on the ($Z,N+1$) isotone following the waiting point by a factor of 100 up or down. 
Such uncertainties of 4 orders of magnitude are typical for reaction rates that are 
dominated by only a few resonances, and where excitation energies are uncertain 
to 100~keV \cite{SBB06}. This is a reasonable range to explore, as here already the reaction 
Q-values are uncertain by at least 100~keV. 
However, it is not an accurate estimate of the uncertainty of statistical model calculations
in the cases under consideration, which is difficult to obtain given the lack 
of experimental and theoretical information. We neglect uncertainties in the $\beta$-decay rates.  
While most $\beta$-decay rates are known with sufficient precision, the partial $\beta$-decay
half-lives of $^{69}$Br, $^{73}$Rb and $^{74}$Sr are not known experimentally. Test 
calculations showed that the results are insensitive to the half-lives of  $^{69}$Br
and $^{73}$Rb because their equilibrium abundances are so low and proton capture tends to 
be faster. Varying the $^{74}$Sr half-life between 10~ms and 100~ms did change the 
$^{72}$Kr effective lifetime by about 10\%.

\begin{figure}
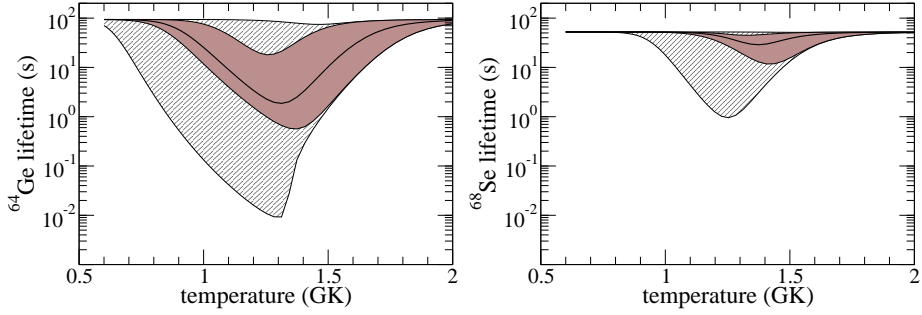

\begin{center}
\includegraphics*[width=6cm]{ge.eps}
\includegraphics*[width=6cm]{se.eps}
\end{center}
\caption{The effective lifetime of $^{64}$Ge (left) and $^{68}$Se (right)
during rp-process conditions as a function of 
temperature taking into acocunt destruction by $\beta$-decay and proton capture. The grey area
denotes the uncertainty due to masses. The hatched area delineates the uncertainty 
when considering an additional error of 
a factor of 100 up or down for the $^{65}$As(p,$\gamma$)$^{66}$Se reaction rate 
in the $^{64}$Ge case and for the $^{69}$Br(p,$\gamma$)$^{70}$Kr rate in the
$^{68}$Se case. }
\label{FigGe}
\end{figure}

\begin{figure}
\begin{center}
\includegraphics*[width=6cm]{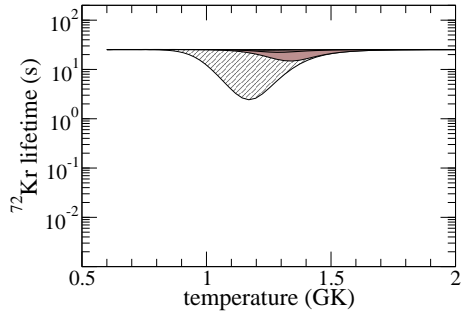}
\end{center}
\caption{Same as Fig.~\protect\ref{FigGe} but for $^{72}$Kr. The reaction rate varied is the 
$^{73}$Rb(p,$\gamma$)$^{74}$Sr rate. This is similar to Fig. 2 in \cite{KR72}.}
\label{FigKr}
\end{figure}

The resulting values and bounds for the effective lifetimes of $^{64}$Ge, $^{68}$Se, 
and $^{72}$Kr for a mass density of $10^6$~g/cm$^3$ and a proton mass fraction of 0.7 are
shown in Figs.~\ref{FigGe} and \ref{FigKr} as functions of temperature. At low temperatures proton captures 
are ineffective. The lifetime is then set by the $\beta$-decay lifetime and is independent of
masses and reaction rates. At the highest temperatures, photodisintegration effectively 
prevents any proton capture flow by driving the abundance distribution of the 
 ($p,\gamma$)-($\gamma,p$) equilibrium towards the waiting point. Again, the 
effective lifetime becomes dominated by the $\beta$-decay lifetime of the waiting 
point nucleus ($Z,N$). The sensitivity of the lifetime to the proton capture rate on the 
isotone following the waiting point ($Z+1,N$) indicates the degree to which equilibrium is established. 
The onset of the rate sensitivity at the lowest temperatures indicates the establishment 
of a  ($p,\gamma$)-($\gamma,p$) equilibrium between the ($N,Z$) and ($N,Z+1$) nuclei. As one can see, such an equilibrium is always established for the temperature 
range of interest here owing to the very low separation energy of the 
($N,Z+1$) nuclei. The disappearance of the reaction rate sensitivity at higher temperatures 
indicates the establishment of full  ($p,\gamma$)-($\gamma,p$) equilibrium between 
the ($Z,N$), ($Z+1,N$) and ($Z+2,N$) nuclei. In that case, the reaction flows become
insensitive to proton capture reaction rates. 

Clearly, in the case of $^{72}$Kr the precision mass measurements of $^{72}$Kr and the 
mirrors to the isotones $^{73}$Rb and $^{74}$Sr, $^{73}$Kr and $^{74}$Kr,
with uncertainties of 8~keV or 
less have drastically reduced mass related lifetime
uncertainties to less than a factor of two. Still, the remaining
$\approx$ 50 \% uncertainty at 1.3~GK might be
relevant for precision tests of X-ray burst models, depending on the lifetimes of the 
preceding waiting points. A further reduction of the uncertainty is however only possible
through a measurement of the mass of $^{73}$Rb, a proton unbound nucleus with a 
lifetime of less than 30~ns \cite{nubase03}. 
Transfer or proton removal reactions might be a possibility to achieve this. 
Reaction rate related uncertainties still allow for the possibility to reduce the 
effective $^{72}$Kr lifetime significantly below the $\beta$-decay lifetime. 

For $^{68}$Se and $^{64}$Ge the situation is worse owing to the larger mass uncertainties
and the higher proton separation energies, which facilitate 2p-captures and increase the 
mass sensitivity of the lifetime. Clearly in these cases the current mass uncertainties
are still too large to allow for reliable X-ray burst simulations.
For example, at 1.4~GK the $^{64}$Ge lifetime ranges from 0.6~s to 40~s
depending on the adopted mass. This is particularly relevant
as this is the first of the major waiting points.
It is important to note that both, the proton separation 
energy of $^{65}$As and the proton separation energy of $^{66}$Se are needed. The uncertainty
in the proton separation energy of $^{65}$As dominates the uncertainty in the lifetime up to
about 1.3~GK. The proton separation energy of $^{66}$Se determines at which temperature 
the $^{66}$Se($\gamma$,p) rate kicks in causing the rise of the lifetime with temperature beyond
about 1.3~GK. Beyond 1.3~GK the uncertainties in the proton separation energies of $^{65}$As and of $^{66}$Se
contribute with roughly equal proportions to the lifetime uncertainty. 

Clearly the current uncertainties in the proton separation energies (see Tab.~\ref{TabMasses})
need to be reduced further by experiments. For the proton separation energy of $^{65}$As,
the 100~keV mass uncertainty of the mirror nucleus $^{65}$Ge contributes significantly. 
A measurement of the $^{65}$Ge mass with an accuracy of at least about 30~keV would therefore 
already reduce the lifetime uncertainty of the $^{64}$Ge waiting point. Such a 
measurement has been performed recently with the ion trap at Michigan State University's
LEBIT facility \cite{Ge65}. An improved $^{70}$Se mass (current uncertainty is 62~keV) 
would slighly improve the calculated $^{70}$Kr mass needed for the $^{70}$Kr proton 
separation energy. Such a measurement should be possible at existing facilities. 

However, as Fig.~\ref{FigGe} shows for the case of $^{68}$Se, an accuracy level of 
the order of 100~keV is not sufficient for reliable rp-process calculations. At least 
for the $^{64}$Ge and $^{68}$Se waiting points a mass accuracy of the order of 10~keV 
would be desirable for the relevant nuclei. 
While there is room for slight improvements in the mass measurements of
$^{64}$Ge (30~keV uncertainty) and $^{68}$Se (20~keV uncertainty), this will not make much
difference until the 100~keV 
error in the Coulomb Shift calculations for the masses of $^{65}$As, $^{66}$Se, 
$^{69}$Br, and $^{70}$Kr is addressed. $^{65}$As, $^{66}$Se and $^{70}$Kr are $\beta$-emitters and 
sufficiently accurate mass measurements at ion traps might become feasible when beam intensities 
at radioactive beam facilities can be improved. The mass of 
proton decaying $^{69}$Br can only be determined through reactions populating 
the short-lived ground state. $\beta$-decay of $^{69}$Kr or proton removal 
reactions are possibilities and might be feasible at existing radioactive beam 
facilities. In parallel, improved theoretical estimates of the 
proton capture rates on $^{65}$As and $^{69}$Br would be helpful, as these
could reduce, or increase, the sensitivities to nuclear masses (see Fig.~\ref{FigGe}).

\section{Conclusions}

Masses play a critical role in the r- and rp-processes. For the rp-process in X-ray bursts, 
masses in the vicinity of the major waiting points $^{64}$Ge, $^{68}$Se, and $^{72}$Kr
are particularly important. While mass uncertainties have been reduced significantly
through experimental and theoretical progress, they are still too large to reliably 
determine the effective lifetimes of these waiting points in the rp-process. Mass
related lifetime
uncertainties for $^{64}$Ge, $^{68}$Se, and $^{72}$Kr still amount to up to
factors of 60, 4, and 1.7 respectively with the combined effective lifetime of all three
waiting points at
1.4~GK ranging from 29~s to 108~s.
In general,
mass accuracies of the order of 10~keV are needed to sufficiently constrain 2p-capture 
flows. While some of the relevant measurements should be feasible at existing 
facilities, others will require considerable advances. 

As we have also shown, 
all lifetimes are sensitive to the proton capture rates on $^{65}$As, 
$^{69}$Br, and $^{73}$Rb. Without reliable estimates for these reaction rates
(upper limits would already be helpful) the question to which degree the 
rp-process reaction flow is impeded by the waiting points $^{64}$Ge, $^{68}$Se, and $^{72}$Kr
cannot be answered reliably. For example, within all the uncertainties the effective 
lifetime of $^{64}$Ge can still range from 10~ms to the full $\beta$-decay lifetime of 92~s. 
In the former case, $^{64}$Ge would not delay the rp-process at all, while in the latter case
it would be the singly most important waiting point imposing a delay of the order of the 
burst timescale. Direct measurements of these reaction rates are difficult, or, in the case
of the proton captures on $^{73}$Rb or $^{69}$Br, impossible due to the sub microsecond
lifetime of the target nuclei. However, decay and transfer reactions clarifying 
the structure of the final nuclei, or Coulomb breakup, might be possibilities in the future. 
Until then, a better theoretical description and further experimental clarification of the 
structure of the mirror nuclei would be helpful.

\section{Acknowledgements}

Support through NSF grants PHY 0110253 and PHY 02 16783 (Joint Institute for Nuclear Astrophysics)
is acknowledged. We thank J. A. Clark for providing his preliminary results for the mass of $^{64}$Ge,
T. Rauscher for providing the NON-SMOKER rates, and F.-K. Thielemann for providing the 
reaction network solver.


\begin{thebibliography}{00}

\bibitem{Kap99} F. K{\"a}ppeler Prog. Part. Nucl. Phys. 43 (1999) 419.

\bibitem{CTT91} J.J. Cowan, F.-K. Thielemann, and J.W. Truran,
Phys. Rep. 208 (1991) 267.

\bibitem{WaW81}
R.~K. Wallace and S.~E. Woosley, Ap. J. Suppl. 45 (1981)  389.

\bibitem{SAG97}
H. Schatz {\it et~al.}, Phys. Rep. 294 (1998) 167.

\bibitem{Fro06} C.~Fr{\"o}hlich {\it et al.} Ap. J. 637 (2006) 415.

\bibitem{Pru05} J. Pruet {\it et al.}, Ap. J. 623 (2005) 325 and astro-ph/0511194.

\bibitem{PKT01} B. Pfeiffer {\it et al.} Nucl. Phys. A 693 (2001) 282.

\bibitem{WGS04} S. Wanajo, S. Goriely, M. Samyn, and N. Itoh, Ap. J. 606 (2004) 1057.

\bibitem{ScR06} H. Schatz and K. E. Rehm, Nucl. Phys. A., to be published

\bibitem{AME03}
  G. Audi, A. H. Wapstra and C. Thibault, Nucl. Phys. A 729 (2003) 129.

\bibitem{AB1}
  B. A. Brown, R. R. C. Clement, H. Schatz, A. Volya, W. A. Richter,
  Phys. Rev. C65 (2002) 045802

\bibitem{Psa04}
  D. Psaltis, astro-ph/0410536 (2004).

\bibitem{StB03}
  T.E. Strohmayer and L. Bildsten, Compact Stellar X-ray Sources, ed.
  W. H. G. Lewin and M. van der Klies (Cambridge: Cambridge Univ. Press),
  astro-ph/0301544 (2003).

\bibitem{KHA99}
O. Koike  {\it et~al.}, Astron. Astrophys. 342 (1999) 464.

\bibitem{SAB01}
H. Schatz {\it et~al.}, Phys. Rev. Lett. 86 (2001)  3471.

\bibitem{WHC04}
S. E. Woosley {\it et al.}, Ap. J. Suppl. 151 (2004) 75.

\bibitem{ICK04}
J. M. M. in't Zandt {\it et al.} astro-ph/0407087 (2004).

\bibitem{Bro04}
E. F. Brown, Ap. J. 614 (2004) 57.

\bibitem{WBS06}
N. N. Weinberg, L. Bildsten, and H. Schatz, Ap. J. to be published
astro-ph/0511247.

\bibitem{FoH64}
W. A. Fowler and F. Hoyle, Ap. J. Suppl. 9 (1964) 201.

\bibitem{SD1}
H. Herndl {\it et al.} Phys. Rev. C 52 (1995) 1078.

\bibitem{FP1}
J. L. Fisker {\it et al.} Atomic Data and Nucl. Data Tab. 79 (2001) 241.

\bibitem{SBB06}
H. Schatz {\it et al.} Phys. Rev. C 72 (2005) 065804.

\bibitem{CLE04}
  R. R. C. Clement {\it et al.}, Phys. Rev. Lett. 92 (2004) 172502

\bibitem{BAB03} K. Blaum, {\it et al.}, Phys. Rev. Lett. 91 (2003) 26.

\bibitem{nubase03}
G. Audi {\it et. al.} Nucl. Phys. A 729 (2003) 3.

\bibitem{PrF03} 
J. Pruet and G. M. Fuller, Astrop. J. Suppl. 149 (2003) 189

\bibitem{CLA04}
  J. A. Clark {\it et al.}, Phys. Rev. Lett. 92 (2004) 192501

\bibitem{GE64} J. A. Clark {\it et al.}, Proceed. of the Fourth Internat. Conf. on
  Exotic Nuclei and Atomic Masses, (2004) 59 and private communication, 
  value from preliminary analysis. 

\bibitem{WAB04} A. W{\"o}hr et al., Nucl. Phys. A 742 (2004) 349. 

\bibitem{KR72}
    D. Rodriguez {\it et al.} Phys. Rev. Lett. 93 (2004) 161104

\bibitem{RaT00} T. Rauscher and F.-K. Thielemann, At. Dat. Nucl. Dat. Tab. 75
(2000) 1. 

\bibitem{Ge65} G. Bollen, et al. private communication 2006. 

\end{thebibliography}
\end{document}